\documentclass [12pt]{article}
\usepackage[utf8]{inputenc}
\usepackage{authblk}
\usepackage{natbib}
\usepackage{babel}
\usepackage{graphicx}
\usepackage{adjustbox}
\usepackage{tabularx}
\usepackage{geometry}
\usepackage{setspace} 
\usepackage{caption}
\usepackage{textcomp}
\usepackage{subcaption}
\usepackage{hyperref}
\hypersetup{
	colorlinks,
	linkcolor={red!50!black},
	citecolor={blue!50!black},
	urlcolor={blue!80!black}
}

\usepackage{tikz}
\usetikzlibrary{
  arrows.meta,
  patterns,
  decorations.pathreplacing,
  calc
}

\usepackage{newtxtext,newtxmath}

\usepackage{amsmath,amssymb}
\usepackage{bm}
\usepackage{longtable}
\usepackage{multicol}
\usepackage{multirow}
\usepackage{booktabs}
\usepackage{makecell}
\usepackage{geometry}
\usepackage{comment}
\usepackage{xcolor}
\usepackage{soul}
\usepackage{pgfplots}
\pgfplotsset{compat=1.17} 
\usepackage{array}
\usepackage{float}

\title{From Tariff Changes to Trade Effects: \\
{\large A Practical Framework for Empirical Trade Policy Analysis}}
\author[1, 3]{Maria Cipollina\thanks{ \ This paper is conceived as a tangible way to honour the memory of Luca Salvatici, a friend and a coauthor that inspired our research and with his attitude and good humour influenced the way in which we leave our professional and personal live. We thank AIEAA and the organizers of the 15th AIEAA Annual Conference, held in Rome and hosted by Universitas Mercatorum, for the opportunity to plan and realize this contribution.}}
\author[2, 3, 4, 5]{Luca De Benedictis}

\affil[1]{\small University of Molise - cipollina@unimol.it;}
\affil[2]{\small Universitas Mercatorum - luca.debenedictis@unimercatorum.it;}
\affil[3]{Centro Rossi-Doria, Univerisità Roma Tre}
\affil[4]{\small Luiss University}
\affil[5]{\small CRIAVET }
\date{}

\begin{document}

\maketitle

\begin{abstract}
After decades of trade liberalization, the 2020s mark a profound shift in the globalization narrative, with tariffs once again becoming a central instrument of trade policy. This reversal also challenges how the effects of trade policy are evaluated. Much of the empirical literature has been developed to assess bilateral tariff reductions or elimination and the associated erosion of preferential advantages. Rising multilateral tariffs pose a different question: they alter not only the trade costs faced by directly affected exporters but also the relative market access of all competing suppliers, potentially reinforcing existing preferences. This paper develops a practical framework for translating tariff changes into trade effects in this new policy environment. We argue that liberalization and protectionism should not be treated as symmetric policy experiments: their effects operate through different changes in relative market access and therefore require careful consideration of how tariff shocks are measured and incorporated into empirical trade models and counterfactual evaluations.
\end{abstract}

\section{Introduction}

For several decades, the evolution of international trade policy was largely framed as a process of progressive liberalization. Successive rounds of multilateral negotiations reduced most-favoured-nation (MFN) tariffs, while bilateral and regional trade agreements increasingly granted selected partners preferential or duty-free access \citep{WTO2026WorldTrade}. Much of the empirical literature therefore developed around policy experiments involving the reduction or elimination of trade barriers, examining how liberalization affects bilateral trade, trade creation and diversion, and welfare \citep{baier2007,Dai2014,Caliendo2015,AndersonYotov2016}.

The trade-policy environment has changed markedly. Since 2018, the United States has introduced substantial tariff increases under Sections 201, 232 and 301, representing the largest rise in US import protection in decades. These measures are part of a broader resurgence of protectionist and strategic trade policies and of an increasingly fragmented international trading system. Recent research shows that their effects extend well beyond reductions in bilateral imports. Higher tariffs affect prices and welfare \citep{Amiti,fajgelbaum2020}, employment both in the imposing country and in third economies \citep{JAVORCIK2025104125,CAVALCANTI2026}, and the organization of global production and sourcing networks \citep{freund2024,kim2025}. A growing literature also documents substantial trade diversion and global reallocation following the US tariff increases, as alternative suppliers respond to changes in the relative access of Chinese products to the US market \citep{Cigna2022,ChoiNguyen2023,Fajgelbaum2024,freund2024, dang2023winners}. Further recent evidence also points to substantial heterogeneity across sectors and production structures \citep{carter2020impact, beames2026} and to broader effects on productivity, resource allocation and structural transformation \citep{kebede2026,kilumelume2026}.

This change in the policy environment also changes the empirical question faced by trade economists \citep{fugazza2013direct}. Much of the existing toolkit was developed in a world in which the relevant experiment was a reduction in bilateral or multilateral trade barriers. Today, researchers are increasingly asked to evaluate the opposite experiment: what happens when MFN tariffs rise? The distinction is not simply one of sign. A bilateral tariff reduction granted to one partner improves that exporter's market access relative to competitors. An increase in an MFN tariff, by contrast, raises trade costs for a potentially broad set of suppliers while leaving preferential tariffs unchanged. The policy change therefore affects not only the countries directly subject to higher tariffs, but also the relative competitive position of exporters whose own tariff treatment has not changed.

This distinction is especially important in a trading system characterized by overlapping preferential agreements. Preferential access is inherently relative: its economic value depends not only on the tariff paid by the beneficiary but also on the tariff treatment faced by competing suppliers. The literature on preferential margins has long emphasized this point \citep{bouet2005,francois2006,Cipollina2011,cipollina2017,cipollina2020}. More recent applications show that changes in MFN tariffs can alter the effective value of existing preferences even when preferential agreements themselves remain unchanged \citep{cipollina2025,Tamberi2026MFN,vanbiesebroeck2026}.

This observation reconnects the current protectionist debate with a long-standing problem in trade-policy analysis: a policy instrument is directly observable, but its economic meaning depends on the benchmark against which it is evaluated. The literature on trade restrictiveness has long shown that theoretically meaningful measures of trade policy require an explicit reference concept rather than a purely statistical aggregation of observed tariffs \citep{AndersonNeary1996,AndersonNeary2003,AndersonNeary2005,cipollina2008}. This distinction is particularly important because simple and trade-weighted tariff averages may provide misleading measures of protection. Since tariffs themselves affect trade flows, highly protected products may receive relatively small weights precisely because protection has reduced their imports \citep{AndersonNeary2005,cipollina2008,Cipollina2011}.

The same problem arises when measuring preferential market access. Comparing a preferential tariff solely with the MFN rate implicitly assumes that the MFN tariff provides the relevant competitive benchmark. Yet exporters compete with suppliers that may themselves enjoy preferential access. When the objective is to measure competitive advantage, the relevant benchmark should reflect the tariff treatment faced by actual competitors, and bilateral market access depends on the entire structure of applied tariffs across suppliers \citep{Cipollina2011}. Building on this insight, \cite{cipollina2017} identify both a \textit{definition bias}, associated with the choice of the reference tariff, and an \textit{aggregation bias}, associated with the construction of that reference across heterogeneous suppliers. Their CES-consistent bilateral tariff margin provides a measure of an exporter's tariff advantage or disadvantage relative to competing suppliers rather than relative to an arbitrary benchmark.

Moreover, structural gravity provides the natural theoretical setting for this relative interpretation of tariff policy. Since \cite{anderson2003}, bilateral trade has been understood as depending on bilateral trade costs relative to the trade costs faced in alternative trading relationships. Subsequent contributions have consolidated structural gravity as the workhorse framework for trade-policy evaluation and counterfactual analysis \citep{head2014,yotov2016,fontagne2022,Yotov2022,Yotov2024}. Recent methodological developments further emphasize the importance of incorporating domestic trade and conducting theory-consistent counterfactual exercises, while flexible structural-gravity approaches allow both liberalization and protection scenarios to be analysed within a common framework \citep{larch2023}.

Yet a common structural framework does not imply that liberalization and protectionism are economically identical experiments. Consider a reduction in the MFN tariff. As non-preferred suppliers obtain better access to the importing market, the relative advantage of preferential partners is eroded. \cite{cipollina2025}, examining the UK Global Tariff, show that MFN liberalization alters not only the tariff paid by affected exporters but also the entire structure of bilateral tariff margins, generating both a direct reduction in protection and an indirect erosion of preferential access. An increase in MFN tariffs operates in the opposite direction: non-preferred suppliers face higher trade costs, while exporters enjoying unchanged preferential tariffs may gain relative market access. Thus, an MFN tariff change can affect countries whose own statutory tariff does not change.

\subsection{The literature space.}
\label{sec:litspace}

\begin{figure}[h!]
    \begin{center}
      \caption{The literature space}\label{fig:literaturespace}\vspace{-0.45cm}
    \includegraphics[trim=1cm 9.2cm 1.5cm 7cm, clip, width=0.90\linewidth]{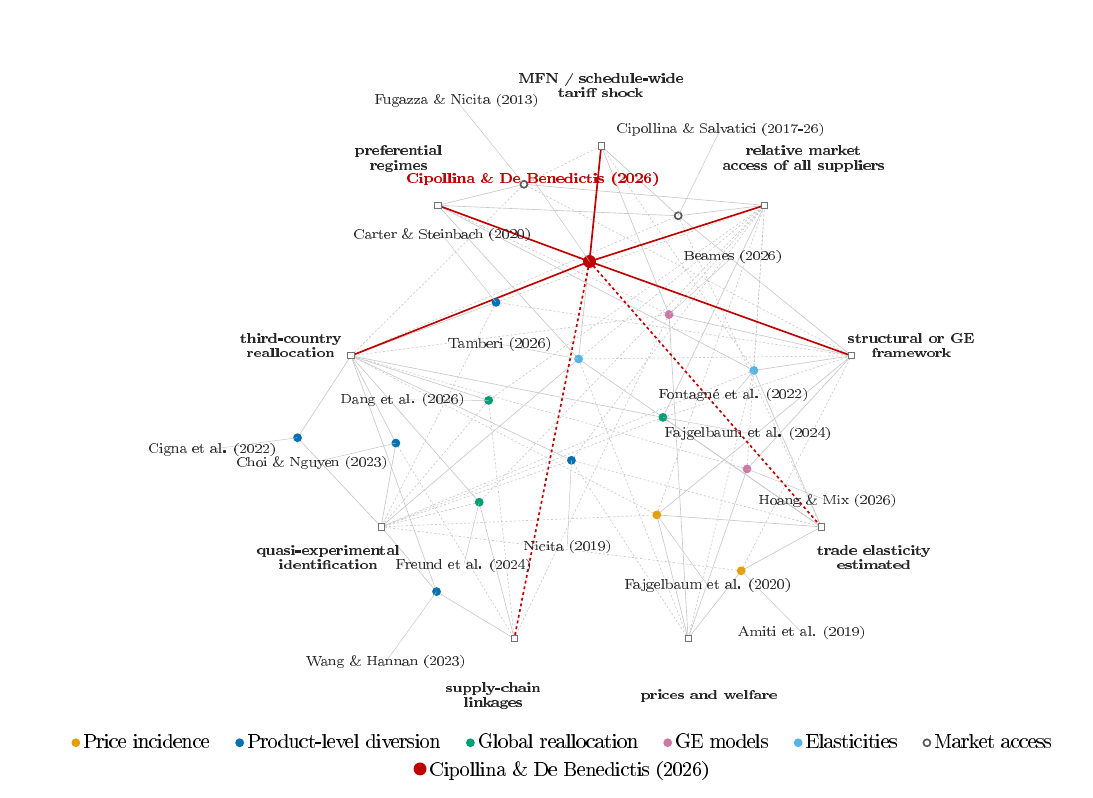} 
    \end{center}
  \begin{footnotesize}
    \begin{spacing}{1}
 \begin{minipage}{\textwidth}
\footnotesize
\emph{Note.}  The figure is a two-mode (affiliation) network of the seventeen studies discussed in this section and of the nine elements on which each of them is coded from its abstract. The elements are fixed on a circle and each study is placed at the barycentre of those it claims, weighted by the strength of the claim, so that proximity reflects the overlap between claims; solid edges denote full claims and dashed edges partial ones. Colours identify the strands of the literature discussed in the text, and the red node is our reference paper \citep{cipollina2026luca}.
\end{minipage}\\\\
      \end{spacing}
      \end{footnotesize}
      \end{figure}

The studies cited above differ along several dimensions at the same time: the tariff variation they exploit, the data they bring to it, the estimator, and the object they ultimately measure. Discussing them one by one makes each difference visible but none of them comparable. We therefore summarise the recent literature on the effects of tariffs as a two-mode network, reported in Figure~\ref{fig:literaturespace}, and built in four steps.

The first step is about selection. The sample consists of the seventeen studies discussed in this section: the empirical literature on the 2018--19 US tariffs and on the trade diversion they generated, the two contributions that extend the question to the 2025--26 tariff wave, the work that treats the tariff elasticity as the parameter of interest, and the methodological antecedents of the relative-market-access approach we adopt.\footnote{\ In the present paper the essays entering the literature review are somehow arbitrary, largely depending on our preferences and subjective knowledge. The first step can be, however, generalized using a systematic review of the literature \citep{lame2019systematic} and eventually a meta-analysis, that can inform the second step of the proposed methodology (see \cite{cipollina2024environmental} for an example).} Working papers are included when they are the current reference for a result, and the three methodological papers by \citet{cipollina2008, cipollina2025} and \citet{cipollina2026luca} are treated as a single entry, since they share the same design.

The second step deals with classification. Each study is summarized on nine elements: an MFN or schedule-wide tariff shock as opposed to a product list aimed at one partner; the relative position of all suppliers; preferential regimes; third-country reallocation; a structural or general-equilibrium framework; a trade elasticity that is estimated rather than assumed; prices and welfare as outcomes; supply-chain linkages; and quasi-experimental identification. The coding is based on the abstract alone. This is a deliberate restriction: the abstract states what the authors themselves regard as their contribution, which makes the classification reproducible and keeps our own reading of each paper out of it. An element takes the value $1$ when the abstract claims it, $0.5$ when the claim is partial or implicit, and $0$ otherwise. Three further elements - a tariff change as the source of variation, product-level trade data, and the United States as the importing market - are shared by almost every abstract and are deliberately left out, since they delimit the field rather than divide it.

The third step is about distance. The procedure in step two returns a $17 \times 9$ matrix whose rows are the studies, and dissimilarity between any two of them is measured by the \citet{gower1971general} distance, the average absolute difference across the nine elements, which ranges from $0$ when two abstracts make identical claims to $1$ when they share none. On this metric the two studies closest to \citet{cipollina2026luca} are \citet{cipollina2017, cipollina2020} at $0.11$ and \citet{fugazza2013direct} at $0.28$, while every other study in the sample lies at $0.33$ or beyond.

Step four is about visualisation. Studies and elements form an affiliation, or two-mode, network \citep{breiger1974duality, borgatti1997network}. The nine elements are fixed on a circle and each study is drawn at the barycentre of the elements it claims, weighted by the strength of the claim, so that studies covering similar sets of elements are placed close to one another; a small repulsion term separates overlapping nodes without altering that ordering. Solid edges mark full claims and dashed edges partial ones, and colours identify the strands of the literature discussed above.

Figure~\ref{fig:literaturespace} shows a literature organised around two poles. The lower-left region collects the reduced-form studies of trade diversion, which measure where displaced imports go but take the tariff shock as a product list and leave the tariff schedule itself unmodelled; the upper and right-hand regions collect the structural and elasticity-based work, which models the schedule but rarely measures reallocation across suppliers. Our paper is the only one in the sample that connects preferential regimes, third-country reallocation and a structural treatment of the tariff schedule at full strength, with five of the nine elements claimed without qualification against a maximum of four elsewhere. It is also, visibly, not connected to quasi-experimental identification, which is precisely where the reduced-form literature is strongest: the exercise describes what each contribution claims, not how well it is executed, and the two poles are better read as complements than as alternatives.\footnote{\ The coding table and the R-code that produces Figure~\ref{fig:literaturespace} are available upon request.}

With the structure of the related literature in mind, this paper, that builds upon \citet{cipollina2026luca}, develops a practical framework for evaluating tariff changes in this renewed protectionist environment. Our purpose is not to propose another index of protection or a new structural trade model. Rather, we ask what is required to translate an observed or counterfactual tariff change into an economically meaningful estimate of its trade effects. We argue that this requires distinguishing between the policy instrument that changes, the resulting change in relative market access, and the trade response generated by that change. These are closely related, but conceptually distinct, objects.

Our contribution is to provide a practical framework that connects the measurement of tariff changes, their implications for relative market access, and the estimation of the resulting trade effects. While the recent literature on US protectionism has provided extensive evidence on direct trade effects and subsequent trade diversion, we focus on a logically prior question: how should the tariff shock itself be represented when the policy change simultaneously alters the relative market access of directly affected and unaffected suppliers? The framework shows why the economic content of a tariff shock depends on the benchmark against which it is measured and on the competitive environment in which it occurs. This issue becomes particularly relevant in a world of overlapping preferential agreements, where a change in MFN protection modifies the relative treatment of suppliers even when many bilateral tariffs remain unchanged. Accordingly, the impact of an MFN tariff increase cannot be inferred from the bilateral tariff change alone, nor can it always be understood as a liberalization exercise with the sign reversed.

The broader contribution is therefore to reconsider the empirical analysis of tariffs at a time when the underlying policy experiment has changed. The methods developed during decades of trade liberalization remain indispensable, but the return of protectionism requires renewed attention to the relative nature of market access and to the mechanisms through which tariff changes propagate across suppliers. Moving from tariff cuts to tariff increases is not merely a change of sign: it changes the configuration of relative market access through which trade adjusts.

The remainder of the paper is organized as follows. Section 2 examines how tariff changes translate into changes in relative market access. Section 3 develops the practical framework for moving from relative market access to trade effects. Section 4 provides an empirical illustration based on counterfactual increases in US tariff protection, examining their implications for bilateral tariff margins and predicted trade reallocation across industries and trading partners. Section 5 concludes.

\section{From Tariff Changes to Market Access}
\label{sec:market_access}

Tariffs are directly observable policy instruments. Their economic interpretation, however, is inherently relative. A change in the tariff applied to one exporter affects not only its own trade cost but also its position relative to other suppliers competing in the same market. Conversely, an exporter may experience a change in relative market access even when the tariff it faces remains unchanged. The relevant policy shock is therefore not fully described by the change in a bilateral tariff alone.

Consider a product \(k\) imported from a set of countries \(i\). Let \(\tau_{ikt}\) denote the tariff applied to exporter \(i\) at time \(t\). The most immediate measure of a tariff reform is the change in the applied bilateral tariff,
\begin{equation}
    \Delta \tau_{ik}=\tau_{ik1}-\tau_{ik0}.
\end{equation}
This measure identifies which bilateral tariffs have changed and by how much, but it does not fully characterize the associated change in market access. Equal changes in statutory tariff rates need not represent equal changes in tariff-inclusive prices. Earlier work on preferential margins addressed this issue by expressing tariff concessions relative to tariff factors rather than simply as percentage-point differences between tariff rates \citep{Cipollina2011}.

More fundamentally, the same bilateral tariff change can have different implications depending on the treatment of competing suppliers. A reduction in the preferential tariff granted to an exporter improves its relative competitive position when competitors' tariffs remain unchanged. By contrast, an increase in the MFN tariff faced by non-preferred suppliers may improve the position of preferential exporters even when their own tariff remains unchanged. For such exporters,
\begin{equation}
    \Delta \tau_{ik}=0,
\end{equation}
although their relative market access changes. Conversely, an MFN tariff reduction may erode existing preferential advantages by improving the competitive position of non-preferred suppliers.

These mechanisms raise the question of the appropriate benchmark against which bilateral tariff treatment should be evaluated. Conventional preference margins compare the preferential tariff with the MFN rate. However, the MFN tariff need not represent the treatment faced by actual competitors, particularly in markets characterized by multiple and overlapping preferential agreements. As emphasized by \citet{Cipollina2011}, measuring competitive advantage requires a benchmark that reflects the tariff treatment prevailing across competing suppliers.

This observation introduces two distinct measurement issues. The first concerns the definition of the reference tariff against which bilateral treatment is evaluated. The second concerns the aggregation of heterogeneous tariff treatments across suppliers. \citet{cipollina2017} formalize these issues as \textit{definition bias} and \textit{aggregation bias}, respectively, and develop a CES-consistent bilateral tariff margin that evaluates an exporter's tariff treatment relative to the competitive conditions prevailing in the importing market.

Under CES preferences, the reference tariff for product \(k\) can be expressed as
\begin{equation}
1+T_{kt}
=
\left[
\sum_i \alpha_i (1+\tau_{ikt})^{1-\sigma_k}
\right]^{\frac{1}{1-\sigma_k}},
\label{eq:ces_reference}
\end{equation}
where \(\sigma_k>1\) denotes the elasticity of substitution across origins and \(\alpha_i\) represents the corresponding normalized CES weight, with \(\sum_i\alpha_i=1\).

The associated bilateral tariff margin is
\begin{equation}
btm_{ikt}
=
\frac{1+T_{kt}}{1+\tau_{ikt}}.
\label{eq:btm_definition}
\end{equation}

The margin measures an exporter's tariff advantage or disadvantage relative to the competitive benchmark. Values above one indicate more favorable tariff treatment than the reference tariff, while values below one indicate a relative disadvantage. Importantly, because the reference tariff depends on the tariff structure across suppliers, a change in competitors' tariffs may alter an exporter's bilateral tariff margin even when its own statutory tariff remains unchanged:
\begin{equation}
\Delta\tau_{ik}=0
\quad\nRightarrow\quad
\Delta btm_{ik}=0.
\label{eq:indirect_market_access}
\end{equation}

The CES formulation also establishes a connection between measurement and behavioral responses. The reference tariff depends on the elasticity of substitution across origins, implying that the construction of relative market access cannot be entirely separated from the parameters governing trade adjustment. This relationship becomes particularly important when estimating trade elasticities and constructing counterfactual policy scenarios, as discussed in the next section.

\subsection{Liberalization, Protectionism and Relative Market Access}
\label{subsec:lib_protection}

The implications of relative market access can be illustrated by comparing two stylized policy experiments: a bilateral tariff reduction and an increase in MFN protection. Both experiments change the tariff treatment of competing suppliers, but they operate in opposite directions for exporters whose own tariffs remain unchanged.

Consider two exporters, \(A\) and \(B\), supplying the same product \(k\). Under the initial tariff structure, their applied tariffs are \(\tau_{Ak}^{0}\) and \(\tau_{Bk}^{0}\), and the corresponding reference tariff is \(T_k^{0}\).

We first consider liberalization. Suppose that the importing country reduces the bilateral tariff applied to exporter \(B\), while leaving the tariff applied to exporter \(A\) unchanged:
\begin{equation}
\tau_{Bk}^{1}<\tau_{Bk}^{0},
\qquad
\tau_{Ak}^{1}=\tau_{Ak}^{0}.
\end{equation}

Holding the other elements of the competitive benchmark constant, the reduction in \(B\)'s tariff lowers the CES reference tariff:
\begin{equation}
T_k^{1}<T_k^{0}.
\end{equation}

Exporter \(B\) benefits directly from the reduction in its own tariff. Exporter \(A\), however, experiences an indirect deterioration in relative market access, despite facing an unchanged statutory tariff. Its bilateral tariff margin changes from
\begin{equation}
btm_{Ak}^{0}
=
\frac{1+T_k^{0}}{1+\tau_{Ak}^{0}}
\qquad \text{to} \qquad
btm_{Ak}^{1}
=
\frac{1+T_k^{1}}{1+\tau_{Ak}^{0}},
\end{equation}
implying
\begin{equation}
T_k^{1}<T_k^{0}
\quad\Rightarrow\quad
btm_{Ak}^{1}<btm_{Ak}^{0}.
\end{equation}

\begin{figure}[htbp]
    \centering
    \includegraphics[width=\textwidth]{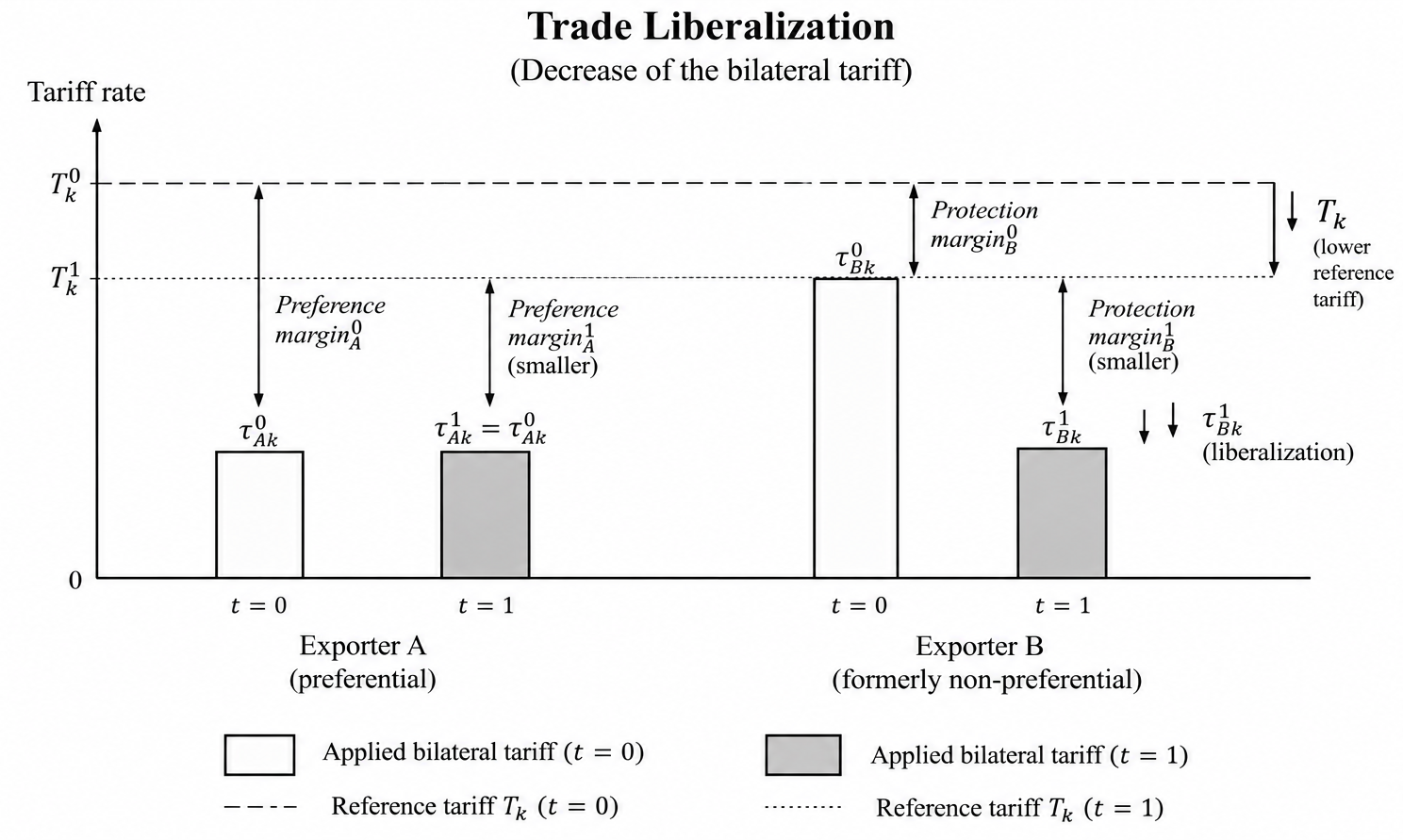}
    \caption{Bilateral tariff liberalization and preference erosion}
    \label{fig:liberalization}
    \begin{minipage}{0.92\textwidth}
    \footnotesize
    \textit{Notes:} The figure illustrates a bilateral tariff reduction for exporter \(B\), while the tariff applied to exporter \(A\) remains unchanged. The reduction in \(B\)'s tariff lowers the reference tariff and consequently reduces exporter \(A\)'s bilateral tariff margin, illustrating preference erosion.
    \end{minipage}
\end{figure}

Figure~\ref{fig:liberalization} illustrates this mechanism. The bilateral liberalization granted to exporter \(B\) reduces the relative advantage enjoyed by exporter \(A\), even though \(\Delta\tau_{Ak}=0\). This is the \textit{preference-erosion} mechanism emphasized in the literature on preferential market access \citep{Cipollina2011,cipollina2017}.

The same framework can be applied to the opposite policy experiment: an increase in MFN protection. Suppose now that exporter \(A\) benefits from preferential treatment, while exporter \(B\) is subject to the MFN tariff. The importing country increases the MFN tariff applied to non-preferential suppliers, leaving the preferential tariff of exporter \(A\) unchanged:
\begin{equation}
\tau_{Bk}^{1}>\tau_{Bk}^{0},
\qquad
\tau_{Ak}^{1}=\tau_{Ak}^{0}.
\end{equation}

The increase in the tariffs faced by non-preferential suppliers raises the CES reference tariff:
\begin{equation}
T_k^{1}>T_k^{0}.
\end{equation}

Although exporter \(A\) experiences no change in its own tariff, its relative market access improves:
\begin{equation}
btm_{Ak}^{0}
=
\frac{1+T_k^{0}}{1+\tau_{Ak}^{0}}
\qquad \text{and} \qquad
btm_{Ak}^{1}
=
\frac{1+T_k^{1}}{1+\tau_{Ak}^{0}},
\end{equation}
so that
\begin{equation}
T_k^{1}>T_k^{0}
\quad\Rightarrow\quad
btm_{Ak}^{1}>btm_{Ak}^{0}.
\end{equation}

\begin{figure}[htbp]
    \centering
    \includegraphics[width=\textwidth]{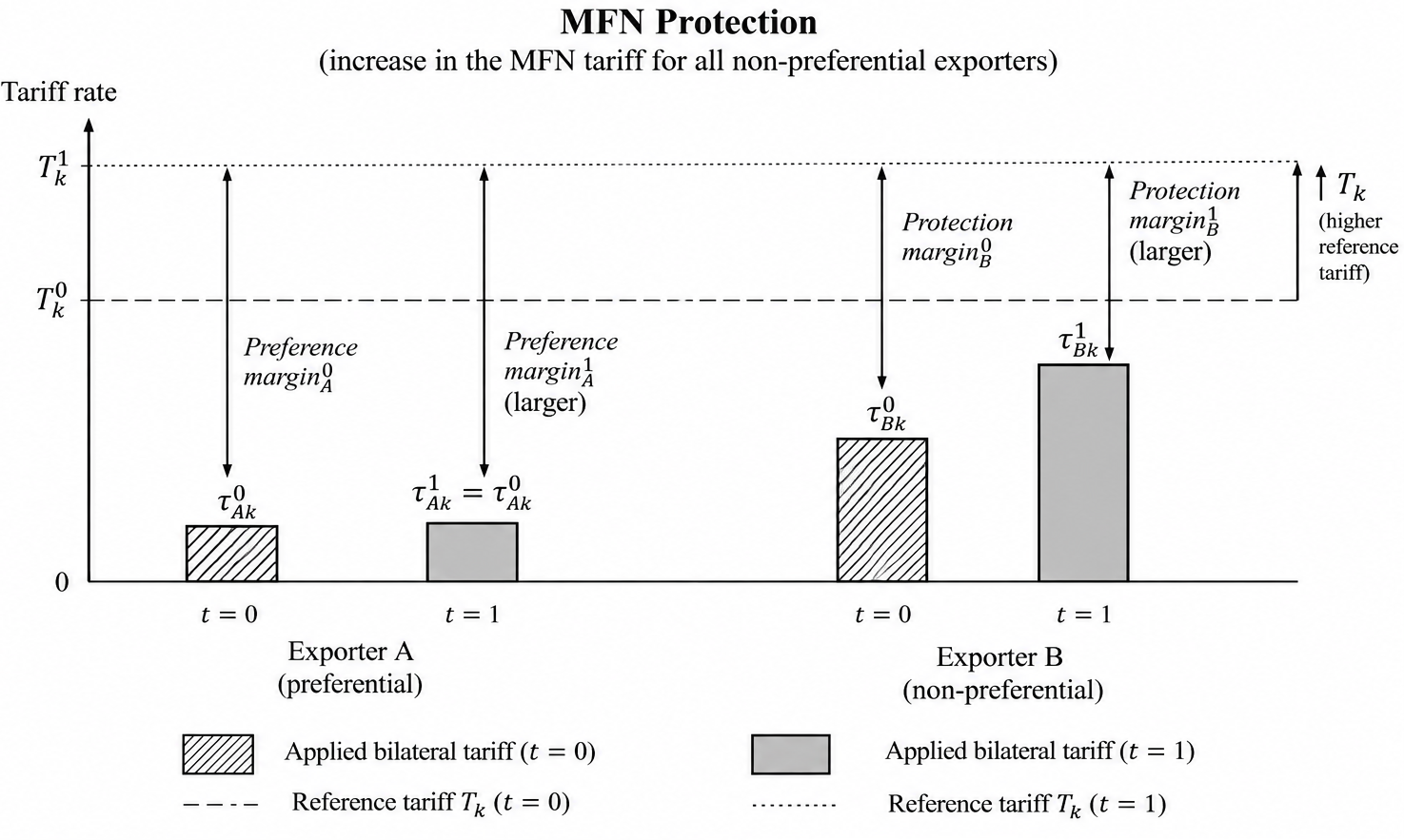}
    \caption{MFN protection and preference amplification}
    \label{fig:protection}
    \begin{minipage}{0.92\textwidth}
    \footnotesize
    \textit{Notes:} The figure illustrates an increase in the MFN tariff faced by non-preferential exporter \(B\), while the preferential tariff applied to exporter \(A\) remains unchanged. The increase in MFN protection raises the reference tariff and consequently increases exporter \(A\)'s bilateral tariff margin, illustrating preference amplification.
    \end{minipage}
\end{figure}

Figure~\ref{fig:protection} illustrates this mechanism. The increase in MFN protection worsens the tariff treatment of non-preferential suppliers, thereby strengthening the relative competitive position of preferential exporters. We refer to this mechanism as \textit{preference amplification}.

The comparison between Figures~\ref{fig:liberalization} and \ref{fig:protection} highlights the two channels through which tariff reforms affect relative market access. The first is a \textit{direct tariff effect}, associated with changes in the tariff applied to a particular exporter. The second is an \textit{indirect competitive effect}, operating through changes in the reference tariff and affecting suppliers whose statutory tariff remains unchanged.

The direction of the indirect effect depends on the policy experiment:
\[
\begin{array}{lll}
\text{Liberalization:} &
\tau_{Bk}\downarrow \Rightarrow T_k\downarrow
\Rightarrow btm_{Ak}\downarrow,
& \text{preference erosion},\\[4pt]
\text{Protection:} &
\tau_{Bk}^{MFN}\uparrow \Rightarrow T_k\uparrow
\Rightarrow btm_{Ak}\uparrow,
& \text{preference amplification}.
\end{array}
\]

This comparison clarifies why protectionism should not be treated simply as liberalization with the sign reversed. Although the underlying measurement framework is common to both experiments, the distribution of tariff changes across suppliers determines which exporters experience direct changes in trade costs and which benefit or lose indirectly through the competitive benchmark.

In particular, an increase in MFN protection may improve the relative market access of preferential exporters without any change in their preferential tariff or in the underlying preferential agreement. A purely bilateral tariff measure would record no policy change for these suppliers, whereas a CES-consistent bilateral tariff margin captures the improvement in their competitive position. This distinction is central to understanding how contemporary tariff increases may generate trade reallocation toward exporters that are not directly targeted by the reform.

\section{From Market Access to Trade Effects}
\label{sec:framework}

Once a tariff reform has been translated into a change in relative market access, the next step is to determine how that change should enter empirical trade-policy analysis. The mapping from tariffs to trade effects is not automatic. It depends on a sequence of choices concerning the definition of the policy experiment, the benchmark against which tariff treatment is evaluated, the aggregation of heterogeneous tariff changes, the behavioral elasticity governing trade responses, and the empirical framework used for estimation and counterfactual analysis.

We organize these choices into four steps. The purpose is not to prescribe a single estimator or a unique policy index, but to make explicit the assumptions required to move from an observed tariff change to an economically interpretable trade effect. A central principle of the framework is that measurement and estimation cannot be treated as independent stages. The way in which the tariff shock is constructed determines the economic variation to which the estimated trade response is attached.

The first step is to define precisely the policy experiment being evaluated. This is often treated as a preliminary descriptive issue, but it is in fact part of the empirical design. A tariff reform may involve a bilateral preferential concession, a change in the MFN schedule, a sector-specific surcharge, or a broader reconfiguration of applied tariffs across groups of exporters. These experiments differ not only in the magnitude of the statutory tariff change, but also in their incidence across suppliers and therefore in the changes in relative market access that they generate.

As illustrated in Section~\ref{sec:market_access}, a bilateral preferential concession and an MFN tariff reform may generate different competitive effects. The former changes the tariff treatment of selected exporters relative to other suppliers, whereas the latter may affect a broader set of exporters while leaving existing preferential tariffs unchanged. Defining the policy experiment therefore requires identifying both the suppliers directly affected by the statutory tariff change and those whose relative market access changes indirectly.

For empirical purposes, the policy experiment should be defined at the most disaggregated level at which tariff treatment differs across exporters and products. Aggregating too early risks obscuring the bilateral variation through which the reform changes competitive conditions. Moreover, conventional statistical aggregation may itself be affected by trade policy. Trade-weighted tariff averages, for example, assign smaller weights to products or suppliers whose imports have already been reduced by high tariffs, potentially understating the intensity of protection \citep{AndersonNeary2005,cipollina2008,Cipollina2011}.

The second step is to select the benchmark against which bilateral tariff treatment is evaluated. As discussed in Section~\ref{sec:market_access}, this choice determines whether the policy variable captures a bilateral statutory tariff change, a conventional preference margin relative to the MFN rate, or a measure of relative market access that incorporates the tariff treatment of competing suppliers.

The choice of benchmark is not merely a statistical aggregation problem. A theoretically meaningful measure of trade policy should be associated with a clearly specified economic reference concept \citep{AndersonNeary1996,AndersonNeary2003,AndersonNeary2005,Cipollina2011}. The MFN tariff provides a simple reference, but it may not adequately represent the competitive environment when several suppliers benefit from preferential treatment. Averages of applied tariffs incorporate a broader set of competitors, although their interpretation depends on the aggregation rule adopted.

When the objective is to evaluate relative market access and its implications for trade volumes, the CES-consistent benchmark introduced in Equation~\eqref{eq:ces_reference} provides a theoretically grounded alternative. The corresponding bilateral tariff margin, defined in Equation~\eqref{eq:btm_definition}, measures an exporter's tariff treatment relative to the competitive benchmark rather than solely relative to its own statutory tariff or the MFN rate.

These alternative measures answer different economic questions and should not be treated as interchangeable. A bilateral tariff change identifies the direct modification of statutory protection. A conventional preference margin measures the advantage relative to the MFN benchmark. A competitor-based average incorporates the tariff treatment of alternative suppliers but depends on the aggregation rule. A CES-consistent bilateral tariff margin additionally accounts for substitution possibilities across origins.

An important implication of the CES formulation is that the reference tariff depends on the elasticity of substitution across suppliers. Consequently, the construction of relative market access cannot be fully separated from the behavioral parameters governing trade adjustment. The elasticity used to construct the tariff benchmark must therefore be consistent with the elasticity used to translate changes in bilateral tariff margins into predicted trade effects. This interdependence motivates the estimation procedure discussed in the following subsection.

 \subsection{Estimate the trade elasticity}
The third step is to estimate how trade flows respond to the policy variable defined in the previous steps. In a structural gravity framework, bilateral trade depends on bilateral trade costs relative to the trade costs prevailing in alternative trading relationships. The interpretation of the estimated elasticity therefore depends critically on how the tariff component of trade costs is represented. When tariffs enter through a bilateral tariff margin, the estimated response refers to changes in relative tariff treatment rather than to changes in the bilateral statutory tariff alone.

A convenient empirical specification is

\begin{equation}
E[M_{ikt}\mid X] = \exp\left[ \beta_k \ln(btm_{ikt}) +\gamma_{it} +\delta_{kt}
\right],
\end{equation}

where \(M_{ikt}\) denotes imports of product \(k\) from exporter \(i\), \(btm_{ikt}\) measures the relative tariff treatment defined in Section~\ref{sec:market_access}, and \(\beta_k\) captures the response of bilateral trade to changes in that relative treatment. The fixed effects control for exporter-time and product-time determinants of trade that are common to the relevant observations.

The appropriate fixed-effect structure depends on the dimensionality of the data and on the policy experiment under investigation. In multi-country structural-gravity applications, importer-time and exporter-time fixed effects can absorb the corresponding multilateral resistance terms, while bilateral fixed effects control for time-invariant bilateral trade costs and unobserved pair heterogeneity \citep{yotov2016,Yotov2022}. In the single-importer setting considered here, exporter-time and product-time fixed effects absorb unobserved determinants of trade that vary along these dimensions. The specification captures conditional trade responses to relative tariff treatment but does not, by itself, constitute a full general-equilibrium structural-gravity model.

PPML provides a natural estimator for this class of models because it accommodates zero trade flows and remains consistent in the presence of heteroskedasticity under standard conditions. More importantly for the present argument, however, the economic interpretation of the estimated coefficient depends on the policy variable entering the gravity equation. If the regressor is the bilateral tariff factor, the coefficient measures the response of trade to changes in bilateral tariff costs. If the regressor is a relative tariff margin, the coefficient measures the response to changes in tariff treatment relative to the competitive benchmark. The same observed tariff reform may therefore generate different empirical variation depending on how the policy shock is represented.

This distinction helps organize the recent empirical literature on tariff increases. Some studies identify trade responses directly from changes in bilateral tariff exposure, exploiting variation across products, origins, or tariff rounds \citep{Amiti,fajgelbaum2020}. Others use treated and untreated products or exporters to identify trade diversion toward third countries \citep{Cigna2022,ChoiNguyen2023,Fajgelbaum2024}, while analyses of global sourcing examine how tariff exposure changes the allocation of imports across suppliers \citep{freund2024}. In these approaches, the response of third-country trade is typically an outcome of the tariff shock. A relative-market-access approach shifts part of this mechanism into the measurement of the shock itself: changes in the tariffs faced by competing suppliers modify the policy variable of exporter \(i\) even when \(\tau_{ikt}\) remains unchanged.

Trade elasticities are also heterogeneous across products and sectors. Product-level evidence indicates substantial variation in the responsiveness of trade to trade costs \citep{fontagne2022}, while recent work suggests that the response to MFN tariff changes may depend on the presence and utilization of preferential treatment \citep{Tamberi2026MFN}. A single elasticity imposed across products may therefore conceal economically important differences in substitution possibilities and in the transmission of tariff shocks.

This issue becomes particularly important when the reference tariff is itself CES-consistent. As established in Section~\ref{sec:market_access}, the reference tariff \(T_{kt}\) depends on the elasticity of substitution across origins, \(\sigma_k\). The construction of the bilateral tariff margin and the behavioral parameter governing trade adjustment are therefore interdependent. This is the empirical counterpart of a more general property of theory-consistent trade-policy indexes: their measurement depends on behavioral responses rather than being separable from them \citep{AndersonNeary2005,Cipollina2011}.

A practical solution is to determine the tariff margin and the relevant substitution parameter iteratively. Starting from an initial benchmark, such as the MFN tariff, an initial bilateral tariff margin is constructed and used to estimate the trade response. The estimated coefficient is then translated into the corresponding substitution elasticity according to the relationship implied by the underlying structural model. This elasticity is used to construct a CES-consistent reference tariff and update the bilateral tariff margins. The trade equation is subsequently re-estimated using the updated margins. The procedure is repeated until convergence is achieved, subject to the existence and stability of a fixed point.

Schematically,
\[
\text{initial benchmark}
\rightarrow
\widehat{\beta}^{(1)}
\rightarrow
\widehat{\sigma}^{(1)}
\rightarrow
T^{CES,(1)}
\rightarrow
btm^{(1)}
\rightarrow
\widehat{\beta}^{(2)}
\rightarrow
\widehat{\sigma}^{(2)}
\rightarrow \cdots .
\]

The exact mapping between \(\beta_k\) and \(\sigma_k\) depends on the underlying structural specification and should not be assumed independently of it. The broader methodological point is that, when the policy variable is constructed using behavioral parameters, measurement and estimation cannot be treated as independent stages. Consistency requires the measure of relative market access and the elasticity used to translate that measure into trade responses to be jointly determined.

\subsection{Counterfactual evaluation}

The final step is to translate the estimated trade response into the effects of a tariff reform. This requires comparing trade under a baseline tariff structure with the trade flows implied by an alternative tariff schedule. The construction of the counterfactual is therefore not independent of the measurement choices made in the previous steps: once market access is defined in relative terms, the relevant tariff structure must be recomputed under the counterfactual policy.

Let \(btm^{0}_{ik}\) denote the bilateral tariff margin under the baseline tariff schedule and \(btm^{1}_{ik}\) the corresponding margin under the counterfactual schedule. Because the benchmark incorporates the tariff treatment of competing suppliers, the counterfactual margin cannot generally be obtained by updating only the bilateral tariffs that change directly. Denoting the baseline and counterfactual reference tariffs by \(T^{0}_{k}\) and \(T^{1}_{k}\), respectively, the relevant margins are

\begin{equation}
btm^{s}_{ik}
=
\frac{1+T^{s}_{k}}{1+\tau^{s}_{ik}},
\qquad s\in\{0,1\}.
\end{equation}

A consistent counterfactual therefore requires recomputing \(T^{1}_{k}\) from the complete counterfactual tariff structure before evaluating the resulting bilateral margins.

As illustrated in Section~\ref{sec:market_access}, a change in the tariffs faced by competing suppliers can alter an exporter's bilateral tariff margin even when its own statutory tariff remains unchanged. A counterfactual based exclusively on bilateral tariff changes would overlook this indirect competitive effect. A relative-market-access counterfactual instead captures both the direct adjustment in tariff treatment and the change in the competitive benchmark generated by the reform.

Once the baseline and counterfactual bilateral tariff margins have been constructed, the estimated trade elasticity can be used to predict the corresponding adjustment in bilateral trade. Drawing on the counterfactual trade-analysis literature, including \citet{lai2004}, the proportional trade response can be expressed as

\begin{equation}
\Delta Trade_{ik}
=
\frac{
\widehat{M}_{ik}(btm^{1}_{ik})
-
\widehat{M}_{ik}(btm^{0}_{ik})
}{
\widehat{M}_{ik}(btm^{0}_{ik})
}.
\end{equation}

Here, \(\widehat{M}_{ik}(btm^{0}_{ik})\) and \(\widehat{M}_{ik}(btm^{1}_{ik})\) denote predicted trade under the baseline and counterfactual tariff structures, respectively, holding the other determinants of trade fixed. This expression measures the proportional change in predicted bilateral trade associated with the reform.

To express the counterfactual response in value terms, the predicted proportional variation can be applied to observed baseline trade flows:

\begin{equation}
TradeEffect_{ik}
=
M^{0}_{ik}
\left(
\frac{
\widehat{M}_{ik}(btm^{1}_{ik})
-
\widehat{M}_{ik}(btm^{0}_{ik})
}{
\widehat{M}_{ik}(btm^{0}_{ik})
}
\right),
\end{equation}

where \(M^{0}_{ik}\) denotes observed baseline trade. This procedure translates the estimated response to changes in relative market access into a counterfactual adjustment expressed in the same units as observed trade. It also distinguishes the proportional response implied by the estimated model from the trade values to which that response is applied.

The counterfactual exercise should distinguish between the theoretical consistency of the tariff measure and the equilibrium closure of the model. Constructing a CES-consistent measure of relative market access ensures that the tariff shock reflects the tariff treatment of competing suppliers, but it does not by itself make the resulting simulation a general-equilibrium exercise. A partial-equilibrium counterfactual can use a CES-consistent bilateral tariff margin while holding expenditures, production, and other determinants of trade fixed. It therefore measures the predicted trade adjustments associated with changes in relative tariff treatment, conditional on the maintained assumptions.

A full structural-gravity counterfactual goes further by allowing the endogenous variables of the model to adjust to the policy shock. Depending on the model closure, this may include changes in expenditures, output, factory-gate prices, and multilateral resistance terms. Partial- and general-equilibrium exercises therefore differ not in whether the tariff shock is measured consistently, but in the set of endogenous adjustments allowed after that shock occurs. Both approaches can be informative, but they answer different policy questions and their estimated trade effects should not be interpreted interchangeably.

The four steps developed above define a practical sequence for empirical tariff analysis:

\[
\text{Tariff shock}
\rightarrow
\text{Market access}
\rightarrow
\text{Elasticity}
\rightarrow
\text{Trade effect}
\]

This sequence makes explicit where alternative empirical approaches may generate different interpretations of the same tariff reform. Differences can arise from the definition of the policy experiment, the choice and aggregation of the tariff benchmark, the elasticity attached to the resulting measure of market access, or the equilibrium closure used in the counterfactual. A bilateral tariff change, an MFN-based preference margin, and a CES-consistent bilateral tariff margin therefore need not imply the same trade response even when they are constructed from the same underlying tariff schedule.

The framework also clarifies how liberalization and protectionism can be analysed within a common empirical structure without being treated as economically symmetric policy experiments. As illustrated in Section~\ref{subsec:lib_protection}, the same measurement principles apply in both cases, but the distribution of tariff changes across suppliers determines the resulting configuration of relative market access. The empirical illustration in the next section applies this framework to a counterfactual increase in US tariff protection, examining how changes in the tariff structure propagate through relative market access into predicted trade adjustments.

\section{From MFN Protection to Trade Reallocation: An Empirical Illustration}

This section illustrates how changes in US tariff protection translate into changes in relative market access and predicted trade flows. Rather than providing a comprehensive evaluation of US trade policy, the exercise demonstrates the empirical implications of measuring tariff protection relative to a CES-based competitive benchmark. We first examine changes in bilateral tariff margins across industries and then consider the corresponding trade adjustments by industry and by exporting country.

The analysis combines bilateral and domestic trade flows from the US International Trade and Production Database for Estimation (ITPDE) with applied and MFN tariffs from WITS--TRAINS. The underlying data cover 2016--2021 and are harmonised across product and industry classifications. Domestic US trade is included because US producers form part of the competitive benchmark faced by foreign suppliers, although domestic transactions are not subject to import tariffs.

The illustrative counterfactual is based on a tariff scenario for February 2026, constructed using information from the Atlantic Council's Trump Tariff Tracker (accessed in May 2026). It introduces an additional 10 percentage points for selected trading partners, 15 percentage points for Bolivia, and sector-specific additions of 25 percentage points for steel and aluminium and for motor vehicles and parts, and 50 percentage points for copper. The sector-specific measures apply to the relevant products across exporters. The scenario represents changes in applied US tariff protection, not necessarily uniform changes in legally defined MFN rates.

Let $\tau_{ikt}^{0}$ and $\tau_{ikt}^{CF}$ denote the initial and counterfactual applied tariffs facing exporter $i$ in product $k$ at time $t$. The counterfactual changes both the exporter's own tariff and the CES reference tariff, $T_{kt}$, calculated from the competitive structure of the US market. Accordingly, the bilateral tariff margin is
\begin{equation}
BTM_{ikt}^{s}=\frac{1+T_{kt}^{s}}{1+\tau_{ikt}^{s}}, \qquad s\in\{0,CF\}.
\label{eq:btm_empirical}
\end{equation}
The change in the logarithm of the margin admits the exact decomposition
\begin{equation}
\Delta\ln BTM_{ikt}=
-\underbrace{\Delta\ln(1+\tau_{ikt})}_{\text{own-tariff component}}
+\underbrace{\Delta\ln(1+T_{kt})}_{\text{competitive-benchmark component}}.
\label{eq:btm_direct_indirect}
\end{equation}
The first component measures the direct change in the tariff applied to exporter $i$; the second captures the indirect adjustment in its relative market access through the tariffs faced by competing suppliers. Thus, an exporter whose own tariff remains unchanged may still experience a change in its bilateral tariff margin.

The substitution elasticities employed in this illustration are recovered from product-level estimates of a Poisson Pseudo-Maximum Likelihood (PPML) gravity specification. These elasticities are used to construct CES-based bilateral tariff margins and predict trade adjustments under the counterfactual scenario.\footnote{The reported results are restricted to products with positive and statistically significant estimated elasticities.}

For each product and supplier, the counterfactual tariff schedule is used to update both the bilateral applied tariff and the CES reference tariff. Predicted trade changes are obtained by comparing the baseline and counterfactual margins using the estimated trade responses and scaling the resulting proportional changes by observed trade. 

\begin{table}[H]
\centering
\caption{Changes in CES-Based Bilateral Tariff Margins by Industry}
\label{tab:btm_scenario1}
\footnotesize
\setlength{\tabcolsep}{5pt}
\renewcommand{\arraystretch}{1.15}

\begin{tabularx}{\textwidth}{
    >{\raggedright\arraybackslash}X
    rrrr
}
\toprule
\textbf{Industry: ISIC 2-digit} &
\textbf{$\sigma$} &
\textbf{$BTM_{CES}$} &
\textbf{$BTM_{SIM}$} &
\textbf{$\Delta BTM$ (\%)} \\
\midrule

Extraction of crude petroleum and natural gas
& 22.78 & 0.995 & 0.984 & -1.11 \\

Manufacture of chemicals and chemical products
& 57.86 & 0.994 & 1.018 & +2.41 \\

Manufacture of coke, refined petroleum products
and nuclear fuel
& 2.78 & 0.981 & 1.002 & +2.14 \\

Manufacture of electrical machinery and apparatus n.e.c.
& 4.91 & 0.999 & 1.027 & +2.80 \\

Manufacture of fabricated metal products
& 30.49 & 0.990 & 0.953 & -3.74 \\

Manufacture of food products and beverages
& 41.90 & 0.982 & 1.022 & +4.07 \\

Manufacture of furniture; manufacturing n.e.c.
& 10.03 & 0.993 & 1.006 & +1.31 \\

Manufacture of machinery and equipment n.e.c.
& 11.99 & 0.989 & 1.006 & +1.72 \\

Manufacture of medical, precision and optical instruments
& 64.33 & 1.015 & 1.095 & +7.88 \\

Manufacture of motor vehicles, trailers and semi-trailers
& 17.18 & 0.990 & 0.960 & -3.03 \\

Manufacture of other non-metallic mineral products
& 7.23 & 0.990 & 1.000 & +1.01 \\

Manufacture of other transport equipment
& 16.65 & 0.970 & 0.981 & +1.13 \\

Manufacture of radio, television and communication equipment
& 41.26 & 0.984 & 1.018 & +3.46 \\

Manufacture of rubber and plastics products
& 46.64 & 0.998 & 1.029 & +3.11 \\

Manufacture of textiles
& 13.23 & 0.993 & 1.008 & +1.51 \\

Manufacture of wood and products of wood and cork
& 4.09 & 0.982 & 0.924 & -5.91 \\

Tanning and dressing of leather; manufacture of
luggage, handbags and footwear
& 5.63 & 1.038 & 1.066 & +2.70 \\

\bottomrule
\end{tabularx}

\vspace{0.2cm}
\begin{minipage}{\textwidth}
\footnotesize
\textit{Notes:} The table reports industry-level mean
elasticities of substitution and CES-based bilateral tariff margins
under the baseline and Scenario 1 tariff schedules.
The percentage change is calculated as
$\Delta BTM = (BTM_{SIM}/BTM_{CES}-1)\times 100$.
Industries assigned the conventional substitution elasticity of two are excluded from the illustration.
Calculations are based on reported industry-level (at ISIC Rev. 3 two-digit industry level) averages.
\end{minipage}
\end{table}

Table~\ref{tab:btm_scenario1} compares the industry-level mean bilateral tariff margins under the initial and counterfactual tariff schedules. The adjustment is heterogeneous: 13 of the 17 reported industries exhibit an increase in the mean margin, whereas four exhibit a decrease. The largest increases arise in medical, precision and optical instruments (7.88\%), food products and beverages (4.07\%), and radio, television and communication equipment (3.46\%). The largest declines occur in wood products ($-5.91\%$), fabricated metal products ($-3.74\%$), and motor vehicles ($-3.03\%$).

These results do not imply that US tariff protection falls in the industries with positive average margins. Rather, a positive change indicates that, on average, the CES competitive benchmark increases relative to the tariff faced by the suppliers represented in the tabulation. The direction of the average margin therefore reflects both own-tariff adjustments and changes in competitors' tariffs, weighted by the underlying market structure and substitution patterns. Because the table compares ratios of industry-level averages, its percentage changes should not be interpreted as average product-level percentage changes or as an explicit numerical decomposition of the two terms in Equation~\eqref{eq:btm_direct_indirect}.

Predicted trade effects are calculated by applying the proportional changes in PPML-predicted trade flows to observed baseline trade, following the counterfactual procedure described in Section~\ref{sec:framework} and the approach discussed by \citet{lai2004}. The resulting adjustments are aggregated by industry and exporting country. The exercise therefore evaluates the implications of changes in applied tariff protection for relative market access and trade reallocation, conditional on the other determinants of trade held fixed in the simulation.

\begin{table}[H]
\centering
\caption{Predicted Trade Effects by Industry}
\label{tab:trade_effects_scenario1}
\scriptsize
\setlength{\tabcolsep}{4pt}
\renewcommand{\arraystretch}{1.15}

\begin{tabularx}{\textwidth}{
>{\raggedright\arraybackslash}X
rrrrr
}
\toprule
& \multicolumn{2}{c}{\textbf{International trade}}
& \multicolumn{2}{c}{\textbf{Domestic US trade}}
& \textbf{Total} \\
\cmidrule(lr){2-3}
\cmidrule(lr){4-5}
\textbf{Industry}
& \textbf{Gains}
& \textbf{Losses}
& \textbf{Gains}
& \textbf{Losses}
& \textbf{Net effect} \\
\midrule

Extraction of crude petroleum and natural gas
& 1,790 & -12,830 & 0 & 0 & -11,040 \\

Manufacture of chemicals and chemical products
& 0 & -12,995 & 0 & 0 & -12,995 \\

Manufacture of coke and refined petroleum products
& 118 & -87 & 0 & 0 & 31 \\

Manufacture of electrical machinery and apparatus
& 1,210 & -194 & 0 & -3,255 & -2,239 \\

Manufacture of fabricated metal products
& 0 & -8,234 & 0 & 0 & -8,234 \\

Manufacture of food products and beverages
& 3,084 & -250 & 0 & 0 & 2,835 \\

Manufacture of furniture; manufacturing n.e.c.
& 197 & -1,559 & 12,766 & 0 & 11,404 \\

Manufacture of machinery and equipment n.e.c.
& 3,812 & -31 & 0 & -10,854 & -7,073 \\

Manufacture of medical, precision and optical instruments
& 1,162 & -515 & 0 & -48,763 & -48,115 \\

Manufacture of motor vehicles, trailers and semi-trailers
& 0 & -74,455 & 0 & 0 & -74,455 \\

Manufacture of other non-metallic mineral products
& 1,191 & -15 & 0 & -6,510 & -5,334 \\

Manufacture of other transport equipment
& 1,834 & -5,470 & 0 & 0 & -3,636 \\

Manufacture of radio, television and communication equipment
& 921 & -28 & 0 & -1,017 & -124 \\

Manufacture of rubber and plastics products
& 1,468 & -4,652 & 0 & 0 & -3,185 \\

Manufacture of textiles
& 27,874 & -219 & 0 & -21,977 & 5,678 \\

Manufacture of wood and products of wood and cork
& 2 & -495 & 1,529 & 0 & 1,036 \\

Tanning and dressing of leather; luggage and footwear
& 937 & -412 & 0 & -881 & -356 \\

\bottomrule
\end{tabularx}

\vspace{0.2cm}
\begin{minipage}{\textwidth}
\footnotesize
\textit{Notes:} Trade effects are expressed in millions of US dollars. The net effect is calculated as the algebraic sum of gains and losses. Totals may differ slightly due to rounding.
\end{minipage}
\end{table}

The sign of an industry's aggregate trade effect need not coincide with the change in its industry-average bilateral tariff margin reported in Table~\ref{tab:btm_scenario1}. Trade effects are calculated at a more disaggregated level and then aggregated, reflecting differences in initial trade values, tariff changes, and substitution elasticities across products and suppliers.

Table~\ref{tab:trade_effects_scenario1} separates positive and negative predicted adjustments for international suppliers and domestic US trade. This is an accounting decomposition by supplier origin, not the direct-versus-indirect decomposition of the tariff margin in Equation~\eqref{eq:btm_direct_indirect}. It illustrates why an analysis restricted to imports may overlook an important part of the adjustment within the US market.

Among foreign suppliers, motor vehicles record particularly large losses, followed by chemicals and petroleum extraction. Textiles instead exhibit sizeable gains. The domestic component may move in the opposite direction: in medical, precision and optical instruments, modest net gains for foreign suppliers coexist with a much larger predicted contraction in domestic trade. A similar offsetting pattern appears in machinery and electrical equipment. Conversely, positive domestic adjustments in furniture and wood products offset losses among international suppliers. These patterns should be interpreted as model-based reallocations of trade flows, not as direct measures of domestic output, employment or welfare.

\begin{table}[H]
\centering
\caption{Predicted Trade Effects for Major US Trading Partners}
\label{tab:country_effects_scenario1}
\small
\setlength{\tabcolsep}{8pt}
\renewcommand{\arraystretch}{1.12}

\begin{tabular}{lrrrr}
\toprule
\textbf{Exporter} &
\textbf{Baseline trade} &
\textbf{Gains} &
\textbf{Losses} &
\textbf{Net change (\%)} \\
\midrule
Canada         & 119,252 & 1,692  & -22,620 & -17.5 \\
Mexico         & 83,738  & 5,630  & -28,070 & -26.8 \\
China          & 61,888  & 16,851 & -6,316  & +17.0 \\
Japan          & 41,257  & 698    & -16,029 & -37.2 \\
Germany        & 28,777  & 704    & -9,203  & -29.5 \\
South Korea    & 16,344  & 939    & -4,962  & -24.6 \\
United Kingdom & 14,583  & 377    & -4,234  & -26.5 \\
Saudi Arabia   & 14,383  & 30     & -1,474  & -10.0 \\
France         & 13,046  & 251    & -2,145  & -14.5 \\
Vietnam        & 12,727  & 604    & -427    & +1.4 \\
Italy          & 12,652  & 735    & -3,238  & -19.8 \\
India          & 8,912   & 4,063  & -1,778  & +25.6 \\
Brazil         & 7,850   & 165    & -760    & -7.6 \\
Iraq           & 7,252   & 0      & -728    & -10.0 \\
Switzerland    & 6,616   & 153    & -1,640  & -22.5 \\
\bottomrule
\end{tabular}

\vspace{0.2cm}
\begin{minipage}{0.95\textwidth}
\footnotesize
\textit{Notes:} Countries are selected according to
baseline predicted trade with the United States.
Gains and losses refer to international trade effects.
The net percentage change is the algebraic sum of
gains and losses divided by baseline predicted trade.
Trade effects are expressed in millions of US dollars. 
\end{minipage}
\end{table}

Table~\ref{tab:country_effects_scenario1} provides a complementary country-level illustration for major US trading partners, selected according to their baseline predicted trade rather than the size of their counterfactual response. Large negative net changes appear for Japan, Mexico, Germany and South Korea, while China and India display positive net changes in the reported aggregation. Vietnam also shows a small positive adjustment. The coexistence of gains and losses within the same exporting country makes the aggregate result sensitive to its product composition, initial trade weights and the sector-specific changes in bilateral tariff margins.

The three tables clarify the sequence from tariff changes to relative market access and then to predicted trade reallocation. The CES benchmark is central because a tariff change affecting one set of suppliers also changes the competitive environment facing others, including domestic producers. The purpose of the exercise is therefore to illustrate the mechanics and empirical relevance of the framework, rather than to rank the winners and losers from US protectionism.

\section{Conclusions}

The recent use of tariffs as instruments of trade policy raises important questions for empirical trade analysis. After decades in which trade-policy evaluation was largely concerned with tariff reductions and preferential liberalization, researchers increasingly face policy experiments involving higher tariffs, differentiated surcharges, and changes in the competitive position of suppliers. Although the empirical tools developed to study liberalization remain essential, their application to protectionist measures requires careful attention to how tariff shocks are defined, measured, and translated into trade effects.

This paper has developed a practical framework connecting tariff changes to relative market access and, ultimately, to predicted trade adjustments. Its central argument is that a change in a bilateral tariff does not fully capture the economic implications of a tariff reform. Market access is inherently relative: exporters compete with suppliers facing potentially different tariff treatments, and changes in the tariffs applied to competitors can affect an exporter's competitive position even when its own tariff remains unchanged. Consequently, the measurement of a tariff shock depends on the benchmark used to evaluate bilateral tariff treatment.

Building on the literature on preferential margins and theory-consistent trade-policy measurement, the framework highlights the role of CES-based bilateral tariff margins in capturing these relative effects. It organizes empirical evaluation into four connected steps: defining the policy experiment, constructing an economically meaningful measure of relative market access, estimating the relevant trade elasticity, and translating the resulting response into counterfactual trade effects. These steps are interdependent. In particular, when the tariff benchmark depends on substitution elasticities, the measurement of market access and the estimation of trade responses must be mutually consistent.

The empirical illustration, based on a counterfactual increase in US tariff protection, demonstrates the relevance of this approach. Changes in applied tariffs generate heterogeneous adjustments in bilateral tariff margins across industries, reflecting both changes in exporters' own tariffs and changes in the competitive benchmark. The corresponding predicted trade effects differ across industries and trading partners, with gains and losses potentially coexisting within the same sector or exporting country. These patterns illustrate why the consequences of tariff increases cannot be inferred from statutory tariff changes alone and why the composition of suppliers and their initial trade positions matter for the resulting reallocation.

The exercise is intended to illustrate the transmission from tariff changes to trade adjustments rather than to provide a comprehensive assessment of US protectionism. The counterfactual results are conditional on the estimated trade responses and the maintained assumptions about other determinants of trade. They should therefore not be interpreted as full general-equilibrium effects or as direct measures of welfare, employment, or production changes. A broader policy evaluation would require modelling the additional adjustments in expenditures, prices, production, and multilateral resistance generated by the reform.

The broader implication is that liberalization and protectionism can be evaluated within a common empirical framework without being treated as economically symmetric experiments. An MFN tariff reduction may erode existing preferential advantages, whereas an increase can reinforce them by worsening the relative position of non-preferred suppliers. More generally, the consequences of any tariff reform depend on how it reshapes the entire structure of relative market access. In an international trading system characterized by overlapping preferential agreements and renewed protectionist pressures, making this relative dimension explicit is essential for producing economically meaningful estimates of trade-policy effects.

\bibliographystyle{apalike}
\bibliography{References.bib}
\end{document}